%% file: ILP_formulations.tex
\documentclass[conference]{IEEEtran}
\IEEEoverridecommandlockouts

\usepackage[utf8]{inputenc}
\usepackage[english]{babel}
\usepackage{csquotes}
\usepackage{standalone}
\usepackage{caption} 
\usepackage{subcaption} 
\usepackage{mathtools} 
\usepackage{amsfonts}
\usepackage{dsfont}
\usepackage{amsthm} 
\usepackage{graphicx}
\usepackage{tikz}
\usepackage{booktabs}
\usepackage[style=ieee, backend=bibtex]{biblatex} 
\usepackage{algorithm}
\usepackage{algpseudocode} 
\graphicspath{ {./images/} }

\newcommand{\N}{\mathbb{N}}
\newcommand{\R}{\mathbb{R}}

\newcommand{\calD}{\mathcal{D}}
\newcommand{\calE}{\mathcal{E}}
\newcommand{\calF}{\mathcal{F}}
\newcommand{\calG}{\mathcal{G}}

\newcommand{\calO}{\mathcal{O}}

\newcommand{\calR}{\mathcal{R}}
\newcommand{\calS}{\mathcal{S}}
\newcommand{\calV}{\mathcal{V}}
\newcommand{\boldalpha}{\boldsymbol{\alpha}}
\newcommand{\boldeta}{\boldsymbol{\eta}}
\newcommand{\boldtheta}{\boldsymbol{\theta}}
\newcommand{\boldlambda}{\boldsymbol{\lambda}}
\newcommand{\boldmu}{\boldsymbol{\mu}}
\newcommand{\boldxi}{\boldsymbol{\xi}}

\newcommand{\boldl}{\mathbf{l}}
\newcommand{\boldv}{\mathbf{v}}
\newcommand{\boldx}{\mathbf{x}}
\newcommand{\boldy}{\mathbf{y}}
\newcommand{\boldz}{\mathbf{z}}
\newcommand{\tildeg}{\tilde{g}}
\newcommand{\tildeh}{\tilde{h}}

\newcommand{\barLambda}{\overline{\Lambda}}
\newcommand{\barmu}{\overline{\mu}}

\newcommand{\eps}{\varepsilon}
\newcommand{\tildePi}{\tilde{\Pi}}
\newcommand{\tildepi}{\tilde{\pi}}

\newcommand{\SOS}{\text{SOS2}}
\newcommand{\minus}{\scalebox{0.4}{$-$}}
\newcommand{\plus}{\scalebox{0.5}{$+$}}

 \def\BibTeX{{\rm B\kern-.05em{\sc i\kern-.025em b}\kern-.08em
 \kern-.1667em\lower.7ex\hbox{E}\kern-.125emX}}

\newtheorem{lemma}{Lemma}[]

\title{Embedding Delay-Constrained VNF Forwarding Graphs into Reconfigurable WDM Optical Networks -- Extended Version \thanks{The authors acknowledge the financial support by the Federal Ministry of Education and Research of Germany in the project “Open6GHub” (grant number: 16KISK011)}}
\author{\IEEEauthorblockN{1\textsuperscript{st} Valentin Kirchner}
\IEEEauthorblockA{\textit{Hasso-Plattner-Institute} \\
\textit{Internet Technologies and Softwarezation}\\
Potsdam, Germany}
\and
\IEEEauthorblockN{2\textsuperscript{nd} Holger Karl}
\IEEEauthorblockA{\textit{Hasso-Plattner-Institute} \\
\textit{Internet Technologies and Softwarezation}\\ 
Potsdam, Germany }}
\date{}

\begin{document}

\maketitle

\begin{abstract}
Operators of reconfigurable wavelength-division multiplexed (WDM) optical networks adapt the lightpath topology to balance load and reduce transmission delays. Such an adaption generally depends on a known or estimated traffic matrix. Network function virtualization (NFV) allows to implicitly change this traffic matrix. However, these two degrees of freedom have largely been considered separately, using resources suboptimally. Especially for delay-sensitive services, an optimal use of resources can be crucial. 

We aim to \emph{jointly} optimize the embedding of virtualized network function (VNF) forwarding graphs with delay constraints and the lightpath topology of WDM optical networks. Unlike previous work, we consider all three types of delays: propagation, processing and forwarding-induced queuing delay. We model the latter two as M/M/1 queues. 

We formulate and analyze a mixed-integer nonlinear program (MINLP), reformulate it as a mixed-integer quadratic constrained program (MIQCP) and approximate it by a mixed-integer linear program (MILP). We evaluate our approach for small-scale examples of a multicast service. \end{abstract}

\section{Introduction}

Classic networks route data over a fixed substrate network topology. The transmission demands can be perceived as the data rate to be provided, averaged over longer time horizons and are usually described by a traffic matrix. Providers route the traffic based on available knowledge of the traffic matrix, service-level agreements and hardware capacities. The options range from fully decentralized, traffic-oblivious solutions \cite{Dalal1978-jl} to more complex, traffic-aware approaches \cite{Cantor1974-qn, Fortz2000-km, Perry2014-ow}.

Besides routing, another degree of freedom is introduced by adjusting the traffic matrix via network function virtualization (NFV). Data flows pass through certain network functions in hardware and software, e.g.\ a firewall, a network address translation device or a deep packet inspector. These functions might change the data rate, e.g. by dropping packets or re-encoding the payload, and take time for processing, using up some of the delay budget of a flow. With virtualization, these functions can be placed and scaled dynamically, changing the traffic patterns in the network. Exploiting this effect has been extensively studied for a \emph{fixed} substrate topology \cite{Agarwal2019-li}.

If a network's substrate allows to reconfigure its topology, it enables a third degree of freedom. An example for such a substrate is an optically switched and wavelength-division multiplexed (WDM) network. It sends data on direct physical lightpaths between potentially non-adjacent vertices. The topology of lightpaths on top of the substrate is the \emph{lightpath topology} and can be optimized for a given traffic matrix. Leveraging this degree of freedom can reduce forwarding delay, energy consumption and capital and operational expenditure by a significant factor compared to static, electrically switched networks \cite{Poutievski2022-tu}.

However, there is a lack of studies that \emph{jointly} exploit flexible VNF placement and scaling \emph{and} the lightpath topology configurations in WDM optical networks. This joint approach can enable delay-critical applications that would otherwise not be possible. We illustrate that with a small example.

\begin{figure*}[h]
  \centering
  \begin{subfigure}{0.3\textwidth}
    \scalebox{0.7}{\input{graphics/motivation1.tex}}
    \caption{Optical network with heterogeneous computing capacities.}  
    \label{motivation_substrate}
  \end{subfigure} \hfill
    \begin{subfigure}{0.3\textwidth}
    \scalebox{0.7}{\input{graphics/motivation2.tex}} 
    \caption{Topology and VNF placement after separate optimization.}
    \label{motivation_standard_topo_reconfigured}
  \end{subfigure}
  \hfill
  \begin{subfigure}{0.3\textwidth}
    \scalebox{0.7}{\input{graphics/motivation3.tex}}
    \caption{Topology and VNF placement after joint optimization.}
    \label{motivation_joint_solution}
  \end{subfigure}
  \caption{Motivation Example}
  \label{fig:motivation_example}
\end{figure*}
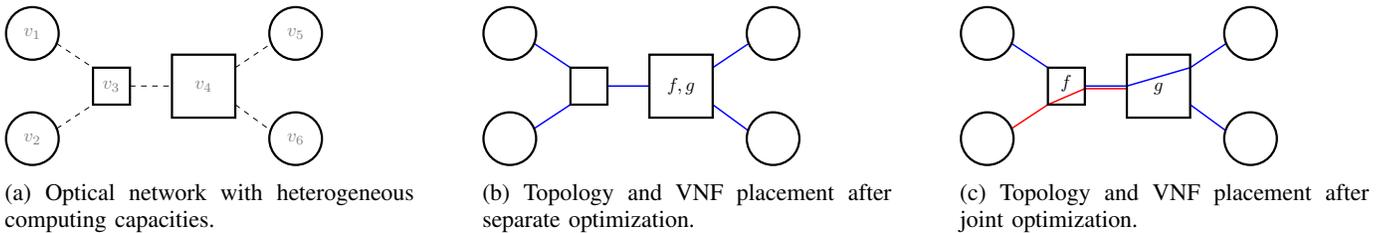

Consider the generic optical network depicted in Figure~\ref{motivation_substrate}. The network offers data transport on two wave lengths (blue and red) over 5 fibers.  We assume that lightpaths between two vertices are bidirectional on exactly one wavelength and that the number of available transceivers per vertex equals its degree. These assumptions guarantee that lightpaths can be established between all physically adjacent vertices, realizing the substrate topology in Figure~\ref{motivation_standard_topo_reconfigured}. Besides data transmission, the network offers the execution of VNFs $f, g$ at vertices $v_3$ and $v_4$. For the simplicity of this particular example, we assume that $f, g$ do not change the data rates. Hardware capacities are often heterogeneous due to an incremental evolution of the network, so we assume that the capacity of $v_4$ is much larger than the capacity of $v_3$. The bigger square around vertex $v_4$ in Figure~\ref{motivation_substrate} illustrates that. 

We consider two flows: The first flow originates at vertex $v_1$ with destination $v_5$ and requests VNF $f$; the second flow originates at vertex $v_2$ with destination $v_6$ and requests VNF $g$. For both flows, we model the arrival of packets as Poisson processes with independent, exponentially distributed packet lengths. We further assume that the time it takes a VNF to process a packet as well as the time it takes a transceiver to forward a packet only depend on the packet length. Hence, we model forwarding at each optical transmitter and processing at each VNF instance as separate M/M/1 queues \cite{kendall}. We use \emph{processing/forwarding delay} to designate the sum of the processing/sending times and the associated queuing times (i.e.\ the sojourn times). In addition, packets experience propagation delay. In total, both flows experience three different delays each: a processing delay, a forwarding delay and a propagation delay. In this simple example, both flows experience a propagation delay independent of the VNF placement and lightpath topology configurations. This allows us to focus our further considerations on the execution and forwarding delay.

First, we optimize the VNF placement for the given substrate topology, as usual. Second, we additionally allow reconfigure the lightpath topology \emph{after} the VNF placement in the substrate topology. Third, we jointly optimize the VNF placement and the lightpath topology.

Given the substrate topology, the placement and scaling of $f$ and $g$ does not change the loads of the lightpaths and only affects the processing delays. Due to the much larger computing capacity at $v_4$, the processing is fastest if both $f$ and $g$ are placed at vertex $v_4$. Given this embedding, we could subsequently optimize the lightpaths. But in the given example, optical transceiver constraints make it impossible to reconfigure the lightpaths. Thus, the first two approaches result in the same topology configuration and VNF placement shown in Figure~\ref{motivation_standard_topo_reconfigured}.
 
As for the third option, the lightpath between $v_3$ and $v_5$ in Figure~\ref{motivation_standard_topo_reconfigured} is potentially highly utilized and queuing can cause significant forwarding delays. If we place $f$ at $v_3$ instead of at $v_4$ we can separate the two flows using the topology depicted in Figure~\ref{motivation_joint_solution}. Although this choice increases the execution delay of $f$, it can drastically decrease the forwarding delay of both flows. Only the joint optimization model can find this optimal tradeoff.

We formulate the joint optimization model as an MINLP, reformulate it as a mixed-integer quadratically constrained program (MIQCP) and approximate it by an mixed-integer linear program (MILP). The problem combines three sub-problems: The placement, scaling and routing for VNF services for a given topology \cite{Draxler2018-hc}; the lightpath topology design for a given traffic matrix \cite{Ramaswami1996-do}; and the routing and wavelength assignment (RWA) for a given lightpath topology \cite{Zang2000-mn}. All three sub-problems are difficult to solve; the nonlinear delay constraints per VNF service requests add even more complexity. This makes even small instances very hard to solve. We reformulate the problem as a non-convex MIQCP and use the highly optimized solver Gurobi \cite{gurobi}, rather than a general purpose MINLP solver. To bound the execution time of the solver, we provide an MILP approximation and evaluate it on small examples.

\section{Related Work}

Researchers extensively studied the problem of configuring optical networks; \textcite{Nance_Hall2021-bs} survey diverse research directions. \textcite{Mukherjee1996-rb} propose an MINLP that is related to our proposed one, but we consider delay constraints for individual services instead of minimizing the mean delay. \textcite{Banerjee2000-oy} study a simplified linear model and provides two heuristics. More recently, \textcite{Jin2016-vd} analyzed optical network configurations in emulations using heuristics; \textcite{Poutievski2022-tu} analyzed optical networks configurations in production using ILP formulations. Both do not consider explicit queuing delays. All of the mentioned studies rely on estimated traffic matrices and cannot be directly transferred to virtual network services. 

The concept of NFV raises the question of optimal resource allocation; \textcite{Gil_Herrera2016-jv} provide an introduction and overview. Our model of NFV requests follows the approach of \textcite{Keller2014-mj} and \textcite{Draxler2018-hc}, where NFV forwarding graphs act as templates and concrete scaling decisions are made during the optimization process. \textcite{Agarwal2019-li} make use of a stricter request model with a fixed forwarding graph; however, they incorporate M/M/1 queues instead of just considering propagation to describe end-to-end delays. We intend to combine the strengths of a more flexible VNF forwarding graph model with the M/M/1 queuing model in the additional context of lightpath topology configuration.

There are only a few publications on joint VNF service provisioning and lightpath topology reconfiguration. All of them are in the context of routing and spectrum allocation (RSA) in elastic optical networks (EON) and focus on EON-related problems like spectrum fragmentation. \textcite{Zeng2016-dk} formulate an exact MILP and compare several heuristics for placement and RSA of tree-type VNF forwarding graphs with depth three. \textcite{Fang2016-dm} and \textcite{Khatiri2022-cg} provide exact ILP formulations and heuristics to embed simple VNF chains. All three papers do not consider delay constraints and work with restrictive VNF forwarding graph models.

\section{Problem Formulation}\label{section_problem_formulation}

\subsection{Remarks on Notation}
For $N \in \N$ with $N \ge 1$ we abbreviate $[N] = \{1, \dots, N\}$ and $[N]_0 = \{0, \dots, N\}$. For a directed graph $\calG = (\calV, \calE)$ we abbreviate an edge $(v, v^\prime) \in \calE$ by $vv^\prime$. For $v \in \calV$ we denote the neighbourhood of incoming and outgoing vertices by $\calV^-(v) = \{ v^\prime \in \calV \; | \; v^\prime v \in \calE \}$ and $\calV^+(v) = \{ v^\prime \in \calV \; | \; v v^\prime \in \calE \}$, respectively. We denote the incoming and outgoing degree by $\deg^-(v) = |\calV^-(v)|$ and $\deg^+(v) = |\calV^+(v)|$, respectively. We denote vectors by bold lowercase letters, e.g. $\mathbf{x} = (x_i)_{i \in [N]} = (x_1, \dots, x_N) \in \R^N$ and write $\mathbf{x} \ge 0$, iff $x_i \ge 0$ for all $i \in [N]$. For a set $A$ we denote its power set by $2^A$. We denote the Dirac function in $y \in \R$ by $\delta_y$, i.e. $\delta_y: \R \to \R$, with $\delta_y(y) = 1$ and $\delta_y(x) = 0$ else.

In the problem formulation below we will use variables with multiple indices. To ease readability, we use lower indices for information regarding the substrate optical network and upper indices for information regarding the network function chains and requests. We hope that there will be no confusion of the upper indices with powers. 

\subsection{Network Model}
We model the substrate optical network as a directed and weighted graph $\calG = (\calV, \calE)$. We refer to $v \in \calV$ as \emph{vertex} and to $e \in \calE$ as \emph{edge}. Every vertex $v$ has a computing capacity $c_v \ge 0$ and can host one or more VNFs $f \in \calF$ at adjustable rate. Furthermore, every vertex $v$ has $\deg(v)$ many transceivers which may be used. Every edge $e$ offers data transmission on a wavelength $\gamma \in \Gamma$ at rate $\barmu \ge 0$ that propagation delay $d_e \ge 0$. We assume that lightpaths are bidirectional and that the network has no wavelength conversion capabilities.

We use the M/M/1 queuing model to approximate the forwarding and VNF execution delay. In particular, we model the arrival of data at the vertices as poisson processes and assume that the execution times of VNFs and the forwarding rates are exponentially distributed. The latter follows from the assumption that packet lengths are exponentially distributed \cite{Fratta1973-nj}. We model the distributions as independent.

\subsection{VNF Forwarding Graph}

We model a VNF forwarding graph as an acyclic, directed and weighted graph $G = (N, A)$ with  $|N| \ge 2$. It has source nodes $S = \{ n \in N \; | \; \deg^-(v) = 0\}$ and destination nodes $D = \{n \in N \; | \; \deg^+(v) = 0\}$. We refer to $n \in N$ as \emph{nodes} and to $a \in A$ as \emph{arcs} to distinguish between the substrate network graph $\calG$ and the VNF forwarding graph $G$. Every non-source and non-destination node $n \in N$ represents a network function $f^n \in \calF$. We denote these functional nodes by $N_\calF = N \setminus (S \cup D)$. 

For $n \in N_\calF$ we approximate its resource consumption $c^n \ge 0$ as affine linearly dependent on the assigned service rate $\mu^n \ge 0$. This takes into account that some VNF are more complex than others. In particular, we set
\begin{align}
c^n = \alpha^n \mu^n + \beta^n, \label{affine_linear_dependency_node}
\end{align}
for $\alpha^n, \beta^n \ge 0$. We approximate the outgoing data rate as affine linearly dependent on the incoming data rates. This takes into account that a VNF can alter the incoming date, e.g. a firewall may drop half of the packets or a server may add advertisement. More precisely, for incoming data rates $\boldlambda^n_{in} = (\lambda^{n^{\minus}n})_{n^{\minus} \in N^{-}(n)}$ we approximate the outgoing data rates $\lambda^{nn^{\plus}} \ge 0$ for $n^{\plus} \in N^+(n)$ by
\begin{align}
\lambda^{nn^{\plus}} = {(\boldalpha^{nn^{\plus}})}^T \boldlambda^n_{in} + \beta^{nn^{\plus}}, \label{affine_linear_dependency_arc}
\end{align}
for $\boldalpha^{nn^{\plus}}, \beta^{nn^{\plus}} \ge 0$.  

Requests of VNF forwarding graphs $G$ for a concrete substrate can restrict the position of nodes $n \in G$. In particular, it is possible to decide the placement and rates of multiple source vertices for one source node, or the placement of multiple destinations for one destination node. Also, it is possible to decide the placement of certain VNFs in the network, e.g. if some of the functionalities need to be processed at specialized hardware boxes is the substrate.

\subsection{Requests}

A network function service request is modeled by a VNF forwarding graph $G = (N,A)$, a maximal flow completion delay $d_{\max} \ge 0$, initial data rates on all outgoing arcs of source nodes specified by $\barLambda \subset \R_+$ and possible restrictions on the positions in the substrate described by $\calS, \calD \subset N \times \calV \times [0,1]$. An element $(n, v, \lambda^n_v) \in \calS$ requires that the proportion of data flowing out of $n$ at $v$ is $\lambda^n_v$. Similarly, $(n, v, \lambda^n_v) \in \calD$ requires that the proportion of data flowing into $n$ at $v$ is $\lambda^n_v$. With the help of these two requirements we can determine, for example, the positions of sources and destinations in the substrate network.

The set of requests is denoted by $\calR$. To reduce the amount of indices, we almost always omit the dependence of $G$ on a particular request $r \in \calR$. 

\subsection{Objectives}
We consider a lexicographic order of four objectives. In the first place, we want to maximize the number of embedded requests that satisfy the delay constraints. We call these \emph{fulfilled requests}, where 'fulfilled' refers to satisfied delay constraints. In the second place, we want to maximize the number of requests that do not satisfy the delay constraints but satisfy the network-capacity constraints. We call those the \emph{unfulfilled but embedded requests}. In that case, we want to minimize the maximal lateness of fullfilled requests. This is our third objective. Our fourth objective is an minimal usage of the network's resources, particularly the accumulated data rates over the lightpaths, the assigned service rates, and the number of established lightpaths.

Our model can easily be adjusted to maximize a notion of profit by assigning a weight to each (un-)fulfilled but embedded requests and maximizing the sum over them. 

\begin{table*}
\centering
\caption{Model Parameters}\label{model_variables}
\begin{tabular}{p{4cm} p{10cm}} \toprule
Parameter & Description \\ \midrule
$\calG = (\calV, \calE)$ & Optical substrate network modeled as a connected, directed and weighted graph. \\
$\Gamma$ & Index set representing available optical wavelengths on the substrate edges. \\
$\calF$ & Set of network functions which can be be executed at the substrate vertices. \\
$c_v \ge 0$ & Computing capacity of a substrate vertex $v \in \calV$.\\
$\barmu \ge 0$ & The data rate carried over a wave length. \\
$d_{e} \ge 0$ & Propagation delay of substrate edge $e \in \calE$. \\
$G = (N, A)$ & NFV Forwarding Graph modeled as acyclic, connected, directed weighted graph. \\
$P$ & Set of all paths in $G$ from source to destination nodes. \\
$S, D, N_\calF \subset N$ & Source, destination and network function nodes of the network function chain.\\
$\barLambda \subset \R_+$ & Initial data rates on the outgoing arcs of source nodes. \\
$\calS, \, \calD \subset N \times \calV \times [0,1]$ &  Set of restrictions of positions in the substrate network. \\
$d_{\max} \ge 0$ & Maximal allowed delay of a flow from source to destination. \\
$\boldalpha, \beta \ge 0$ & Affine linear parameters, compare (\ref{affine_linear_dependency_node}) and (\ref{affine_linear_dependency_arc}). \\
$\calR$ & Set of requests. \\\bottomrule
\end{tabular}
\end{table*}

\section{MINLP Formulation}\label{section_nlp}

\subsection{Variables}
The main decision variables are $\boldl, \, \boldlambda$ and $\boldmu$. The variable $\boldl \in \{0, 1\}$ decides the lightpath topology, $\boldlambda \ge 0$ the embedding of VNF forwarding-graph arcs and $\boldmu \ge 0$ the service rates of VNFs. The variables $\boldx, \, \boldy$ and $\boldz$ are derived from the main variables. Here, $\boldx = (x_1, x_2, x_3, x_4)$ holds information on the status of requests (fulfilledness, embeddedness, lateness, maximal lateness), $\boldy$ on the VNF placement and $\boldz$ on the usage of lightpaths. All variables except $\boldl$ and $x_4$ depend on a particular request $r \in \calR$, but we omit the upper index $r$ for readablility. 

More precisely, $l_{w, w^\prime, e, \gamma} \in \{0,1\}$ decides if a lightpath is set up from vertex $w$ to $w^\prime$ over the edge $e \in \calE$ on wavelength $\gamma$. The variable $\lambda^a_{v, v^\prime, w, w^\prime} \ge 0 $ decides if and how much data is routed from vertex $v$ to $v^\prime$ over a lightpath from vertex $w$ to $w^\prime$ to serve arc $a$. In particular, this means for $n^{\minus} \in N^-(n)$ if $\lambda^{n^{\minus}n}_{v, v^\prime, w, w^\prime} > 0$, then the node $n^{\minus}$ is placed at vertex $v$ and node $n$ is placed at vertex $v^\prime$. The vertices $w$ and $w^\prime$ determine the route from $v$ to $v^\prime$. This is necessary as we might not be able to establish a direct lightpath from $v$ to $v^\prime$. The assigned service rate to $n \in N_{\calF}$ at $v \in \calV$ is decided by $\mu^n_v \ge 0$. The binary variables $x_1, \, x_2 \in\{0,1\}$ indicate if a request is fullfilled embedded or just embedded, i.e. $x_1 = 1$ implies $x_2= 1$. The lateness of a request is denoted by $x_3 \ge 0$, the maximal lateness of all requests by $x_4 \ge 0$. The auxiliary variables $y^n_v, z^a_{v, v^\prime, w, w^\prime} \in \{0,1\}$ indicate if an instance of VNF $n$ is placed on vertex $v$ and if $\lambda^a_{v, v^\prime, w, w^\prime} > 0$, respectively.

\begin{table*}
\centering
\caption{Decision Variables}\label{Variables}
\begin{tabular}{p{4cm} p{10cm}} \toprule
Variable & Description \\ \midrule
$\lambda^a_{v, v^\prime, w, w^\prime} \ge 0$ & Data rate from $v \in \calV$ to $v^\prime \in \calV$ over lightpath from $w \in \calV$ to $w^\prime \in \calV$ to serve $a \in A$.  \\
$\mu^n_v \ge 0$ & Service rate assigned to $n \in N_\calF$ at $v \in \calV$. \\
$l_{w,w^\prime, e, \gamma} \in \{0,1\}$ & 1 iff a lightpath is set up from $w \in \calV$ to $w^\prime \in \calV$ over edge $e \in \calE$ on $\gamma \in \Gamma$. \\
$x_1 \in \{0,1\}$ & 1 iff the request is fulfilled. \\ 
$x_2 \in \{0,1\}$ & 1 iff the request is unfulfilled but embedded. \\ 
$x_3, \, x_4 \ge 0$ & Lateness of the request and maximal lateness of all requests. \\ 
$y^n_v \in \{0,1\}$ & 1 iff $n \in N$ is placed at $v \in \calV$. \\
$z^a_{v,v^\prime, w, w^\prime} \in \{0,1\}$ & 1 iff $\lambda^{a}_{v, v^\prime, w, w^\prime} > 0$. \\ \bottomrule
\end{tabular}
\end{table*}

\subsection{Constraints}
Our model allows the decision whether or not a request is embedded and where the VNFs should be served. That means that the (multi-class) traffic matrix describing demands between substrate vertices is variable. The affine linear dependency between incoming and outgoing data rates further complicates dependencies. Also, the logical topology is allowed to be a complete directed graph with undirected self-loops: the last two lower indices $w, \, w^\prime$ of $\lambda^a_{v, v^\prime, w, w^\prime}$ can be thought of as an edge in the logical topology. This generality makes the formulation of flow conservation rules more subtle than standard formulations.

Recall that we suppress dependencies on requests $r \in \calR$ for readability. The following constraints hold for all $r \in \calR$.

\subsubsection{Auxiliary Variables}
\begin{flalign}
&& x_1 &\le x_2 \label{obj_1_to_obj_2} \\
&& x_3 &\le M \cdot \big(1 - x_1 \big) \label{obj_3_to_obj_1} \\
&& x_3 &\le x_4 \label{max_lateness} \\
\shortintertext{$\forall v \in \calV, \, n \in N_\calF, \, n^{\minus} \in N^-(n):$}
&& \sum_{v^\prime, w^\prime \in \calV} \lambda^{n^{\minus}n}_{v^\prime, v, w^\prime, v} &\le M \cdot y^n_v \label{placement_constr1} \\
&& y^n_v &\le M \cdot \sum_{v^\prime, w^\prime \in \calV} \lambda^{n^{\minus}n}_{ v^\prime, v, w^\prime, v} \label{placement_constr2} \\
\shortintertext{$\forall a \in A, \, v, v^\prime, w, w^\prime \in \calV:$}
&& \lambda^a_{v, v^\prime, w, w^\prime} &\le M \cdot z^a_{v, v^\prime, w, w^\prime} \label{z_activation1} \\
&& z^a_{v, v^\prime, w, w^\prime}  &\le M \cdot \lambda^a_{v, v^\prime, w, w^\prime}. \label{z_activation2} 
\end{flalign}
$M > 0$ denotes a large enough constant. Constraints (\ref{obj_1_to_obj_2}) and (\ref{obj_3_to_obj_1}) ensure that a request cannot be fulfilled if it is not embedded or its lateness is greater than zero. Constraint (\ref{max_lateness}) defines $x_4$ as the maximal lateness over all requests. Constraints (\ref{placement_constr1}) and (\ref{placement_constr2}) activate the binary auxiliary variable $y^n_v$ if vertex $v $ is the endpoint of any route serving arc $n^{\minus}n$. The variable $\boldy$ is needed to add constant terms in (\ref{vertex_util}), (\ref{highlevel_flow_constr}) and to compute the execution delay in (\ref{exec_del}). Constraint (\ref{z_activation1}) and (\ref{z_activation2}) are used to activate the auxiliary variable $\boldz$. This variable is used to prohibit multiple paths in (\ref{no_path_multiplexing}) and to compute the propagation and forwarding delay in (\ref{prop_del}), (\ref{forw_del}).

\subsubsection{Placement and Routing Constraints}

\begin{flalign}
\shortintertext{$\forall \, s \in S, \, n \in N^{+}(s), \, \overline{\lambda^{ns}} \in \barLambda:$}
&& \overline{\lambda^{ns}} \cdot x_2 =& \sum_{v, v^\prime, w^\prime \in \calV} \lambda^{sn}_{v, v^\prime, v, w^\prime} \label{start_constr} \\
\shortintertext{$\forall \, (n, v, \lambda^n_v) \in \calS:$}
& \mathrlap{\lambda^n_v \, \sum_{n^{\plus} \in N^+(n)} \, \sum_{v^\prime, v^{\prime \prime}, w^\prime \in \calV} \, \lambda^{nn^{\plus}}_{v^{\prime \prime}, v^\prime, v^{\prime \prime}, w^\prime}}  && \nonumber \\ 
&& =& \sum_{n^{\plus} \in N^+(n)} \, \sum_{v^\prime, w \in \calV} \lambda^{nn^{\plus}}_{v, v^\prime, v, w^\prime} \label{equ:source_pos_constr} \\
\shortintertext{$\forall \, (n, v, \lambda^n_v) \in \calD:$}
& \mathrlap{\lambda^n_v \, \sum_{n^{\minus} \in N^-(n)} \, \sum_{v^\prime, v^{\prime \prime}, w \in \calV} \, \lambda^{n^{\minus}n}_{v^{\prime \prime}, v^\prime, w, v^\prime}} && \nonumber \\
&& =&  \sum_{n^{\minus} \in N^-(n)} \, \sum_{v^{\prime \prime}, w \in \calV} \lambda^{n^{\minus}n}_{v^{\prime \prime}, v, w, v} \label{equ:dest_pos_constr} \\
\shortintertext{$\forall \, n \in N \setminus S, \, n^{\plus} \in N^+(n), \, v \in \calV:$}
& \mathrlap{\sum_{n^{\minus} \in N^-(n)} \, \sum_{v^\prime, w^\prime \in \calV} \alpha^{nn^{\plus}}_{n^{\minus}n} \cdot \lambda^{n^{\minus}n}_{v^\prime, v, w^\prime, v} + \beta^{nn^{\plus}} \cdot y^n_v} && \nonumber \\ 
&& =& \sum_{v^\prime, w^\prime \in \calV} \, \lambda^{nn^{\plus}}_{v, v^\prime, v, w^\prime} \label{highlevel_flow_constr} \\
\shortintertext{$\forall a \in A, \, v,v^\prime \in \calV, \, w \in \calV \setminus \{v, v^\prime\}:$}
&& 0 =& \sum_{w^\prime \in\calV} \lambda^a_{v, v^\prime, w^\prime, w} - \sum_{w^\prime \in \calV} \lambda^a_{v, v^\prime, w, w^\prime} \label{midlevel_flow_constr}	 \\
\shortintertext{$\forall a \in A, \, v,v^\prime, w \in \calV:$}
&& \sum_{w^\prime} z^a_{v, v^\prime, w, w^\prime} & \le 1 \label{no_path_multiplexing} \\
\shortintertext{$\forall a \in A, \, v, v^\prime, w \in \calV, \, \text{s.t. } v \neq w \lor v^\prime \neq w:$}
&& \lambda^a_{v, v^\prime, w, w} &= 0 \label{equ:no_cycle_1} \\
&& \lambda^a_{w, w, v, v^\prime} &= 0  \label{equ:no_cycle_2} \\
\shortintertext{$\forall a \in A, \, v, v^\prime, w \in \calV, \, \text{s.t. } v \neq v^\prime :$} 
&&  \lambda^a_{v, v^\prime, w, v} &= 0  \label{equ:no_cycle_3} \\
&& \lambda^a_{v, v^\prime, v^\prime, w.} &= 0 \label{equ:no_cycle_4} 
\end{flalign}
Constraint (\ref{start_constr}) sets the initial data rate of embedded requests according to $\barLambda$. Constraints (\ref{equ:source_pos_constr}) and (\ref{equ:dest_pos_constr}) restrict the placement of nodes on vertices according to $\calS$ and $\calD$. Constraint (\ref{highlevel_flow_constr}) establishes the affine linear relation between incoming and outgoing data rates. But this constraint has two more purposes. On the one hand, it ensures the correct embedding of destination demands specified by $(n^{\plus}, v_d) \in \calD$ if $n^+$ is a destination node. On the other hand, it is a flow conservation rule at the level of network function placements: it ensures that the endpoint $v$ of an embedded network function chain arc $n^{\minus}n$ is the start point of the next network function chain arc $nn^{\plus}$. Constraint (\ref{midlevel_flow_constr}) is a flow conservation constraint. To obtain a unique delay on a route from $v$ to $v^\prime$ serving arc $a$, constraint (\ref{no_path_multiplexing}) enforces that this route is unique. The Constraints (\ref{equ:no_cycle_1})-(\ref{equ:no_cycle_4}) prohibit useless self-loops and ensures that sources cannot be sinks and vice versa.

\subsubsection{Capacity Constraints}

\begin{flalign}
\shortintertext{$\forall v \in \calV:$}
&& \sum_{r \in \calR} \, \sum_{n \in N_\calF} \alpha^n \cdot \mu^n_v + \beta^n \cdot y^n_v &\le   c_v \label{vertex_util} \\
\shortintertext{$\forall v \in \calV, \, n \in N_\calF:$}
&& \sum_{n^{\minus} \in N^-(n)} \, \sum_{v^\prime, w \in \calV} \lambda^{n^{\minus}n}_{v^\prime, v, w, v} &\le \mu^n_v \label{capacity_constr2} \\
\shortintertext{$\forall w, w^\prime \in \calV,\, w \neq w^\prime:$}
&&  \sum_{r \in \calR} \, \sum_{a \in A} \, \sum_{\substack{v, v^\prime \in \calV \\ v \neq v^\prime}}  \lambda^a_{v, v^\prime, w, w^\prime} & \le  \sum_{\gamma \in \Gamma} \, \sum_{u \in \calV^+(w)} \barmu \cdot l_{w, w^\prime,  wu, \gamma}. \label{lightpath_activation_with_bitrate}
\end{flalign}
Constraint (\ref{vertex_util}) has two purposes. Firstly, it establishes the affine linear connection between the assigned service rates and the used computing resources at the substrate network vertices. Secondly, it ensures that the computational capacity of the vertices is not exceeded. Constraint (\ref{capacity_constr2}) guarantees that the assigned service rate on $v$ to $n$ suffices. Constraint (\ref{lightpath_activation_with_bitrate}) connects the two decision variables $\boldlambda$ and $\boldl$: if variable $\lambda^a_{v, v^\prime, w, w^\prime} > 0$ indicates the usage of a lightpath from vertex $w$ to $w^\prime$, then the lightpath has to be specified in terms of variables $l_{w, w^\prime, e, \gamma}$. In particular, the lightpath has to start on an edge adjacent to $w$. Furthermore, the total traffic over a lightpath should not exceed the maximal data rate $\barmu$.

\subsubsection{Optical Constraints}

\begin{flalign}
\shortintertext{$\forall w, w^\prime \in \calV, \, u \in \calV \setminus \{w, w^\prime\}, \, \gamma \in \Gamma:$}
&& \mathllap{\sum_{u^\prime \in \calV^-(u)} l_{w, w^\prime, u^\prime u, \gamma}} &= \sum_{u^\prime \in \calV^+(u)} l_{w, w^\prime, u u^\prime, \gamma} \label{lightpath_flow} \\
\shortintertext{$\forall e \in \calE, \, \gamma \in \Gamma:$}
&& \sum_{w, w^\prime \in \calV} l_{w, w^\prime, e, \gamma} &\le 1 \label{lightpath_unique_wavelength} \\
\shortintertext{$\forall w, \, w^\prime \in \calV, \, e \in \calE:$}
&& \sum_{\gamma \in \Gamma} l_{w, w^\prime, e, \gamma} &\le 1 \label{one_lightpath} \\
\shortintertext{$\forall w, \, w^\prime \in \calV, \, uu^\prime \in \calE, \, \gamma \in \Gamma :$}
&& l_{w, w^\prime, uu^\prime, \gamma} &= l_{w^\prime, w, u^\prime u, \gamma} \label{bidirectional lightpaths} \\
\shortintertext{$\forall w \in \calV :$}
&& \sum_{w^\prime \in \calV} \sum_{u \in \calV^+(w)} \sum_{\gamma \in \Gamma} l_{w, w^\prime, wu, \gamma} & \le  \deg (w) \label{transceiver_constr} \\
\shortintertext{$\forall w \in \calV, \, e \in \calE \, \gamma \in \Gamma:$}
&& \mathllap{l_{w, w, e, \gamma}} &= 0 \label{equ:no_light_cycle_1} \\
\shortintertext{$\forall w, w^\prime \in \calV, \, u \in \calV^-(w), \, u^\prime \in \calV^+(w^\prime) \, \gamma \in \Gamma:$}
&& l_{w, w^\prime, uw, \gamma} &= 0 \label{equ:no_light_cycle_2} \\
&& l_{w, w^\prime, w^\prime u, \gamma} &= 0 \label{equ:no_light_cycle_3} 
\end{flalign} 
Constraint (\ref{lightpath_flow}) is a flow conservation constraint on the level of wavelengths. Constraint (\ref{lightpath_unique_wavelength}) ensures that distinct lightpaths use distinct wavelengths over the same edge. Constraint (\ref{one_lightpath}) restricts a lightpath to one wavelength. Constraint (\ref{bidirectional lightpaths}) ensures that lightpaths are bidirectional. Constraint (\ref{transceiver_constr}) makes sure that the number of transceivers is less than the node degree. Similar to Constraints (\ref{equ:no_cycle_1})-(\ref{equ:no_cycle_4}) the constraints (\ref{equ:no_light_cycle_1})-\ref{equ:no_light_cycle_3}) prohibit useless cycles.

\subsubsection{Delay Constraints}

\begin{flalign}
\shortintertext{$\forall (n)_{j \in [J]} \in P, \, \boldv \in \calV^J:$}
&& &\sum_{j \in [J-1]} \, \sum_{w, w^\prime \in \calV} z^{n_j n_{j+1}}_{v_j, v_{j+1}, w, w^\prime} \cdot \psi_{w, w^\prime} \label{prop_del} \\
&& +\; & \sum_{j \in [J-1]} \sum_{\substack{w, w^\prime \in \calV \\ w \neq w^\prime}} z^{n_j n_{j+1}}_{v_j, v_{j+1}, w, w^\prime} \cdot \varphi_{w, w^\prime} && \label{forw_del} \\
&& +\; & \sum_{j \in [J-2]} \, y^{n_{j+1}}_{v_{j+1}} \cdot \rho^{n_{j+1}}_{v_{j+1}} \le  d_{\max} + x_3, \label{exec_del}
\end{flalign} 
where
\begin{align*}
	\psi_{w, w^\prime} &:= \sum_{e \in \calE} \, \sum_{\gamma \in \Gamma} d_e \cdot l_{w, w^\prime, e, \gamma} \\
	\varphi_{w, w^\prime} &:= \Big(\barmu - \sum_{r\in \calR} \, \sum_{a \in A} \, \sum_{v, v^\prime \in \calV } \lambda^a_{v, v^\prime, w, w^\prime} \Big)^{-1} \\
	\rho^n_v &:= \Big( \mu^{n}_{v} -  \sum_{n^{\minus} \in N^-(n)} \, \sum_{v^\prime, w \in \calV} \lambda^{n^{\minus}n}_{v^\prime, v, w, v} \Big)^{-1}.
\end{align*}

There might be multiple paths $p \in P$ from source to destination nodes in $G$. As we allow to scale out each VNF, we need to bound the flow completion delay of any possible embedding of $p$ into $\calG$. This is formalized in (\ref{prop_del}) - (\ref{exec_del}). More precisely, the constraint ensures that the total delay of \emph{any} flow on the path $(n_1, \dots, n_J) \in P$ embedded onto the vertices $(v_1, \dots, v_J) \in \calV^J$ does not exceed the sum of the delay bound $d_{\max}$ and the lateness $x_3$. The total delay of such a flow is given by the sum of the propagation, forwarding and execution delays. The propagation delay (\ref{prop_del}) is the sum of lightpath propagation delays. The forwarding (\ref{forw_del}) and execution (\ref{exec_del}) delays are given by the sojourn time of a M/M/1 queue.

\subsection{Objective Function}

We split the overall objective function into four partial objective functions with lexicographic order. Our main objective is to maximize the number of fulfilled requests
\begin{align*}
o_1 = \sum_{r \in \calR} x^r_1.
\end{align*}
The secondary objective is to maximize the number of fulfilled and unfulfilled but embedded requests
\begin{align*}
o_2 = \sum_{r \in \calR} x^r_2.
\end{align*}
Third, we aim to minimize the maximal lateness of all embedded requests
\begin{align*}
o_3 = x_4.
\end{align*}
Our fourth objective is to minimize the weighted sum of the number of established lightpaths weighted by their propagation delay, the accumulated data rates over the configured substrate network and the accumulated service rates of the VNFs 
\begin{align*}
o_4 = c_1 \,  o_{\text{path}} + c_2 \, o_{\text{data}} + c_3 \, o_{\text{proc}},
\end{align*}
where $c_1, c_2, c_3 \ge 0$. Here, we set
\begin{align}
o_{\text{path}} &= \sum_{w, w^\prime \in \calV} \sum_{e \in \calE} \sum_{\gamma \in \Gamma} d_e \cdot l_{w, w^\prime, e, \gamma} \nonumber \\
o_{\text{data}} &= \sum_{r \in \calR} \, \sum_{a \in A} \Bigg( \sum_{\substack{v, v^\prime, \\ w, w^\prime \in \calV}} \lambda^a_{v, v^\prime, w, w^\prime} - \frac{1}{2} \sum_{v \in \calV} \lambda^a_{v, v, v, v} \Bigg) \nonumber \\
o_{\text{proc}} &= \sum_{n \in N_{\calF}} \, \sum_{v \in \calV} \mu^n_v \nonumber
\end{align}
In the definition of $o_{\text{data}}$, we scale down variables of the form $\lambda^a_{v,v, v, v}$ by $\frac{1}{2}$ as they correspond to embeddings of arcs that do not generate any traffic in the substrate network. The total objective is the minimization of
\begin{align*}
o = C_3o_3 + C_4o_4 -C_1o_1 - C_2o_2
\end{align*}
for $C_i \ge 0, \, i\in [4]$. Here, $C_i$ should be choosen such that the $o_i$ satisfy a lexicographic order with $o_{i+1} \le o_i$.

\subsection{Complexity} \label{section:complexity}
First, we look at the space complexity of the problem. We denote the maximal number of arcs in any NFV Forwarding Graph request by $m = \max_{\calR}|A|$ and obtain that the size of variable $\boldlambda$ is in $\calO(m|\calR||\calV|^4)$. This is not surprising, as we need to decide for any request $r \in \calR$ and arc $a \in A$ a start and endpoint $v, v^\prime \in \calV$ and a route in the complete graph over $\calV$ with $2|\calV|^2$ edges. The size of variable $\boldl$ is in $\calO(|\calV|^2 |\calE| |\Gamma|)$, as we need to decide for any lightpath between $w, w^\prime \in \calV$ the routing and wavelength assignment. If we drop the Delay constraints (\ref{prop_del})-(\ref{exec_del}) and assume that $|\Gamma| \le m|\calR|$ (more wavelengths than arcs to embed are unnecessary), we obtain a number of constraints in $\calO(m|\calR||\calV|^3)$. The delay constraints strongly depend on the length $j$ of the longest path from any source to any destination node in a requested VNF forwarding graph. Denoting $p = \max_\calR |P|$ we then obtain that the delay constraints are in $\calO(p \calR |\calV|^j)$. 

Since the problem contains two NP-complete problems as a sub-problem (placement, scaling and routing of VNFs \cite{Draxler2018-hc} and RWA \cite{Mukherjee1996-rb}) it is NP-complete.

\subsection{MIQCP Reformulation}\label{section:MIQCP}

\begin{figure*}[h]
	\centering
	\begin{subfigure}{0.3\textwidth} \centering
		\includegraphics[scale=0.27]{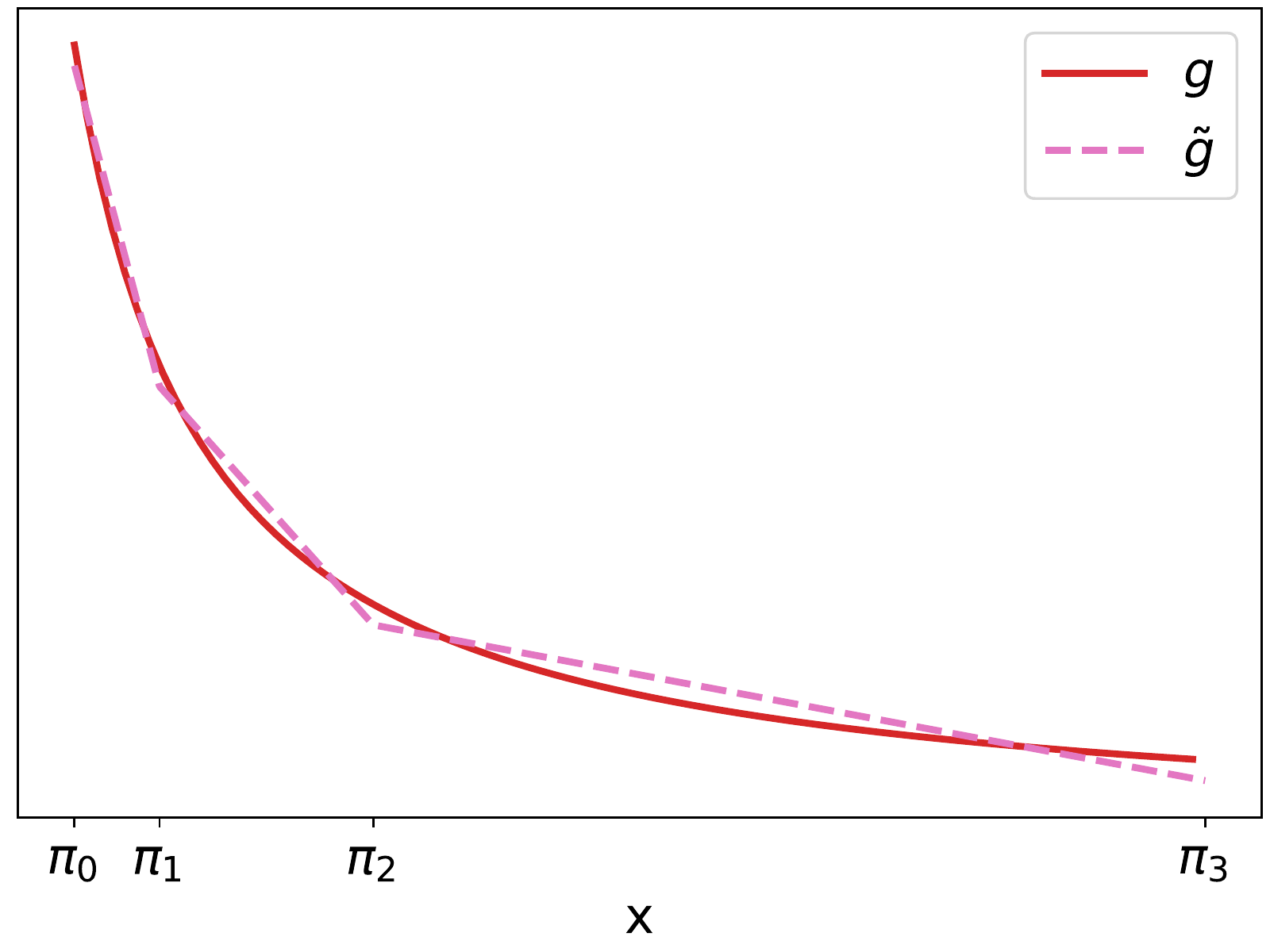}
		\caption{Approximation of $g$ by $\tildeg$.}
		\label{fig:approx_g}
	\end{subfigure} \hfill
	\begin{subfigure}{0.3\textwidth} \centering
		\includegraphics[scale=0.27]{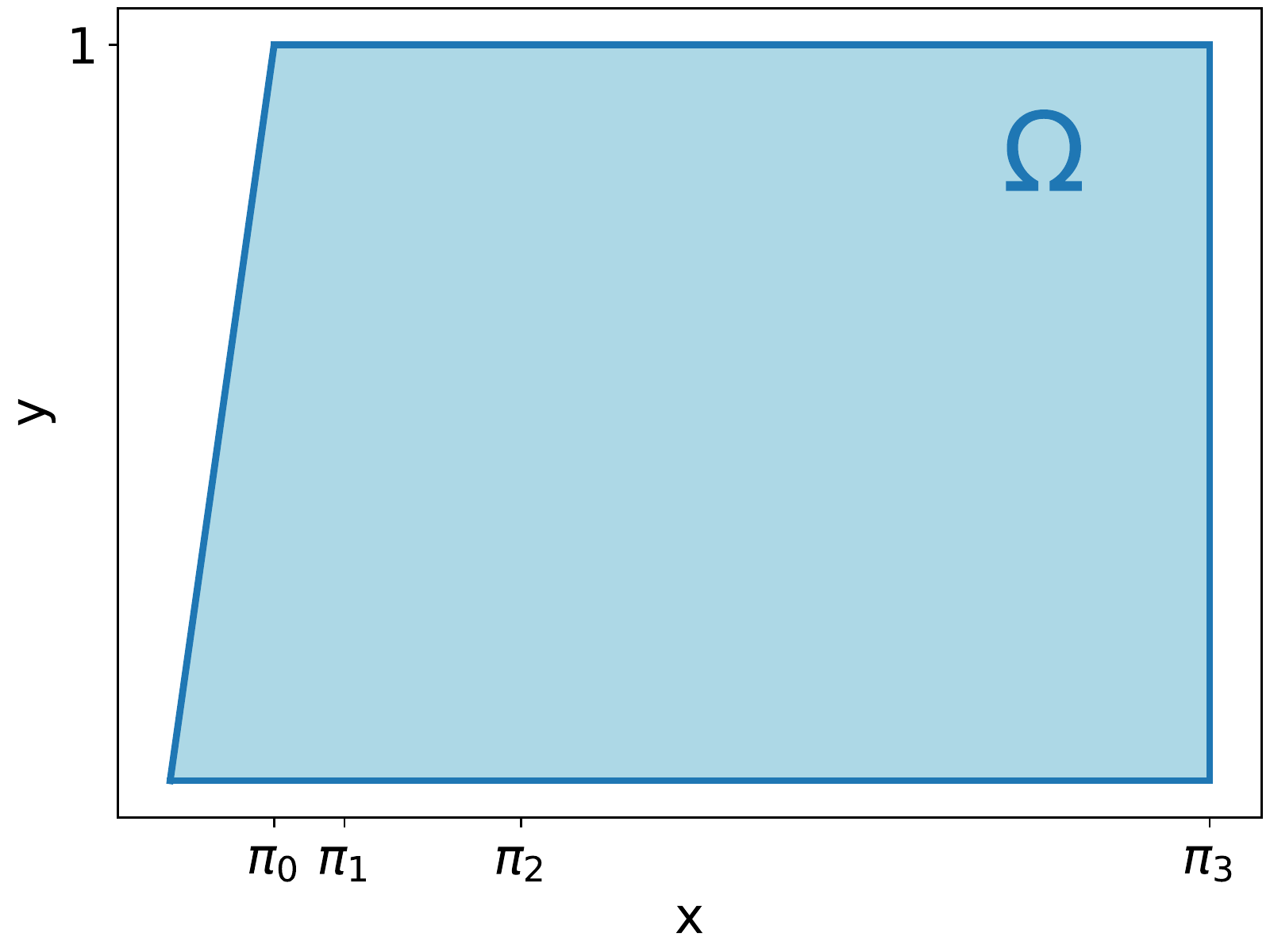}
		\caption{Domain $\Omega$ of $h$.}
		\label{fig:domain_omega}
	\end{subfigure} \hfill
	\begin{subfigure}{0.3\textwidth} \centering
		\includegraphics[scale=0.18]{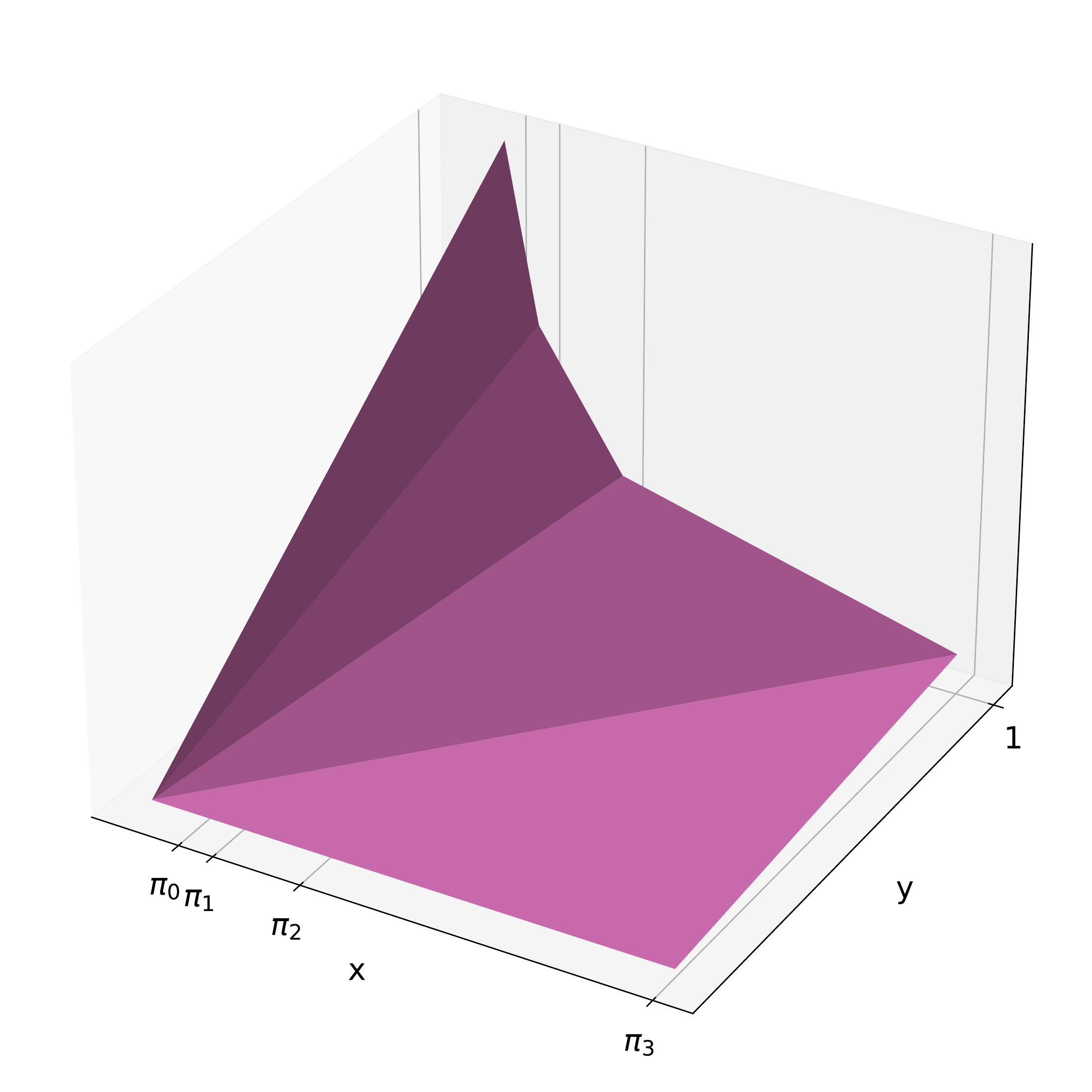}
		\caption{Approximation $\tildeh$.}
		\label{fig:approx_h}
	\end{subfigure}
	\caption{Illustration of the Approximation.}
\end{figure*}

As the MINLP has a large space and time complexity, even small instances are difficult to handle with general mixed-integer, non-linear and non-convex solvers. Therefore, we reformulate the problem as a non-convex MIQCP for which specialized solvers like Gurobi \cite{gurobi} are available.

\begin{lemma} (MIQCP Reformulation)
	We can reformulate the MINLP as an equivalent MIQCP with the same space complexity up to a constant factor. 	
\end{lemma}
\begin{proof}
We replace the rational terms of the form $(\mu - \lambda)^{-1}$ in the delay constraints (\ref{forw_del}), (\ref{exec_del}) by auxiliary variables: for (\ref{forw_del}), we introduce the auxiliary variables $\boldeta$ with constraints
\begin{flalign*}
\shortintertext{$\forall w, w^\prime \in \calV, \, w \neq w^\prime:$}
	&& 0 & \le \eta_{w, w^\prime}\\
	&& 1 + \sum_{r\in \calR} \, \sum_{a \in A} \, \sum_{v, v^\prime \in \calV} \lambda^a_{v, v^\prime, w, w^\prime} \cdot \eta_{w, w^\prime} &\le \barmu \cdot \eta_{w, w^\prime},
\end{flalign*}
for (\ref{exec_del}) we introduce the auxiliary variables $\boldtheta$ with constraints
\begin{flalign*}
\shortintertext{$\forall n \in N_{\calF}, \, v \in \calV:$}
	&& 0 & \le \theta^n_v \\
	&& y^n_v + \sum_{n^{\minus} \in N^-(n)} \, \sum_{v^\prime, w \in \calV} \lambda^{n^{\minus}n}_{v^\prime, v, w, v} \cdot \theta^n_v &\le \mu^n_v \cdot \theta^n_v.
\end{flalign*}
We can then rewrite the delay constraints (\ref{prop_del})-(\ref{exec_del}) as quadratic constraints:
\begin{flalign*}
\shortintertext{$\forall (n)_{j \in [J]} \in P, \, \boldv \in \calV^J:$}
&& & \sum_{j \in [J-1]} \, \sum_{w, w^\prime \in \calV} \, \sum_{e \in \calE} \, \sum_{\gamma \in \Gamma} d_e \cdot  z^{n_j n_{j+1}}_{v_j, v_{j+1}, w, w^\prime} \cdot l_{w, w^\prime, e, \gamma}  \\
&& + & \sum_{j \in [J-1]} \sum_{\substack{w, w^\prime \in \calV \\ w \neq w^\prime}} z^{n_j n_{j+1}}_{v_j, v_{j+1}, w, w^\prime} \cdot \eta_{w, w^\prime} \\
&& + & \sum_{j \in [J-2]} \, y^{n_{j+1}}_{v_{j+1}} \cdot \theta^{v_{j+1}}_{n_{j+1}} \le d_{\max} + x_3.
\end{flalign*}
The number of added variables and constraints does not alter the space-complexity classes determined in Section~\ref{section:complexity}.
\end{proof} 

\section{MILP Approximation}\label{section:approx}

A significant part of the problem's time complexity arises from the nonlinear terms in the delay constraints (\ref{prop_del})-(\ref{exec_del}). We propose two steps to reduce this complexity. First, we only allow to establish lightpaths on a fixed given route, e.g. the shortest path with respect to the propagation delay. The quadratic expression in (\ref{prop_del}) then simplifies to a linear one. Second, we approximate the nonlinear terms in (\ref{forw_del}) and (\ref{exec_del}) by piecewise-linear functions. In total we obtain an MILP formulation that approximates the exact MINLP formulation.

\subsection{Wavelength Assignment Only}
We see two possibilities for restricting the route on which lightpaths can be established: we could either extend the above model by additional constraints or reduce it accordingly. For completeness, we state our reduction in Appendix \ref{appendix:WA only}.

\begin{table*}[h!]
\centering
\caption{Evaluation Setup}\label{table:eval_setup}
\begin{tabular}{p{4.2cm} p{10cm}}
\toprule
Parameter & Description \\ 
\midrule
$\calG = (\calV, \calE)$ & See Figure~\ref{fig:topos}.\\
$\Gamma = [6]$ & The number of available wavelengths is chosen such that each connection from source to destination can potentially use a dedicated wavelength for the hop from source to computing vertex and for the hop from computing vertex to destination.  \\
$\calF = \{f\}$ & Generic VNF. \\
$c_v \in \{5, 50\}$ & Heterogeneous computing capacities. \\
$\barmu = 4$ & Sending rate of the optical transceivers. \\
$d_{e} = 0.1$ & Propagation delay on each edge. \\
$G = (N, A)$ & \scalebox{0.8}{\input{graphics/simple_nfc_tikz.tex}} \\
$\calS = \{(s, v, 1)\}$ & Source position: $s$ is positioned on vertex $v$ with proportion $1$. \\
$\calD = \{(d, v_i, \frac{1}{3})\}_{i=1,2,3}$ & Destination positions: $d$ is positioned at nodes $v_i$, each with proportion $\frac{1}{3}$.  \\
$\barLambda = \{\lambda^s\} = \{3\} $ & Initial aggregated source data rate at $s$.\\
$d_{\max} = 0$ & The requested delay bound is set to zero as we are interested in the minimal delay bound that can be fulfilled. \\
$\boldalpha = \boldsymbol{1}, \, \beta = \boldsymbol{0}$ & Chosen for simplicity. \\
$|\calR| = 1$ & One multicast request. \\
$C_2 = 100$, $C_3 = 10$, $C_1=C_4=0$ & Our primary goal is to embed the request, our secondary goal is to minimize its lateness. \\
\bottomrule
\end{tabular}
\end{table*}

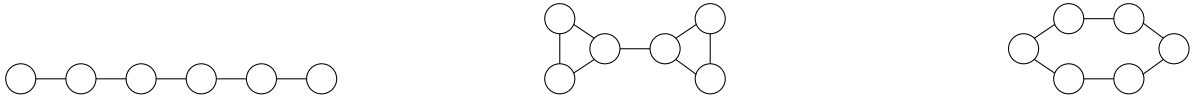
\begin{figure*}[h!] \centering
	\begin{subfigure}{0.32\textwidth} \centering
		\scalebox{1.0}{\input{graphics/path_graph.tex}}
	\end{subfigure} \hfill
	\begin{subfigure}{0.32\textwidth} \centering
		\scalebox{1.0}{\input{graphics/barbell_graph.tex}}
	\end{subfigure} \hfill
	\begin{subfigure}{0.32\textwidth} \centering
		\scalebox{1.0}{\input{graphics/ring_graph.tex}}
	\end{subfigure}
	\caption{Substrate Topologies}
	\label{fig:topos}
\end{figure*}

\begin{figure*}[h!]
 \centering
  	\begin{subfigure}{0.32\textwidth} \centering
    \includegraphics[scale=0.3]{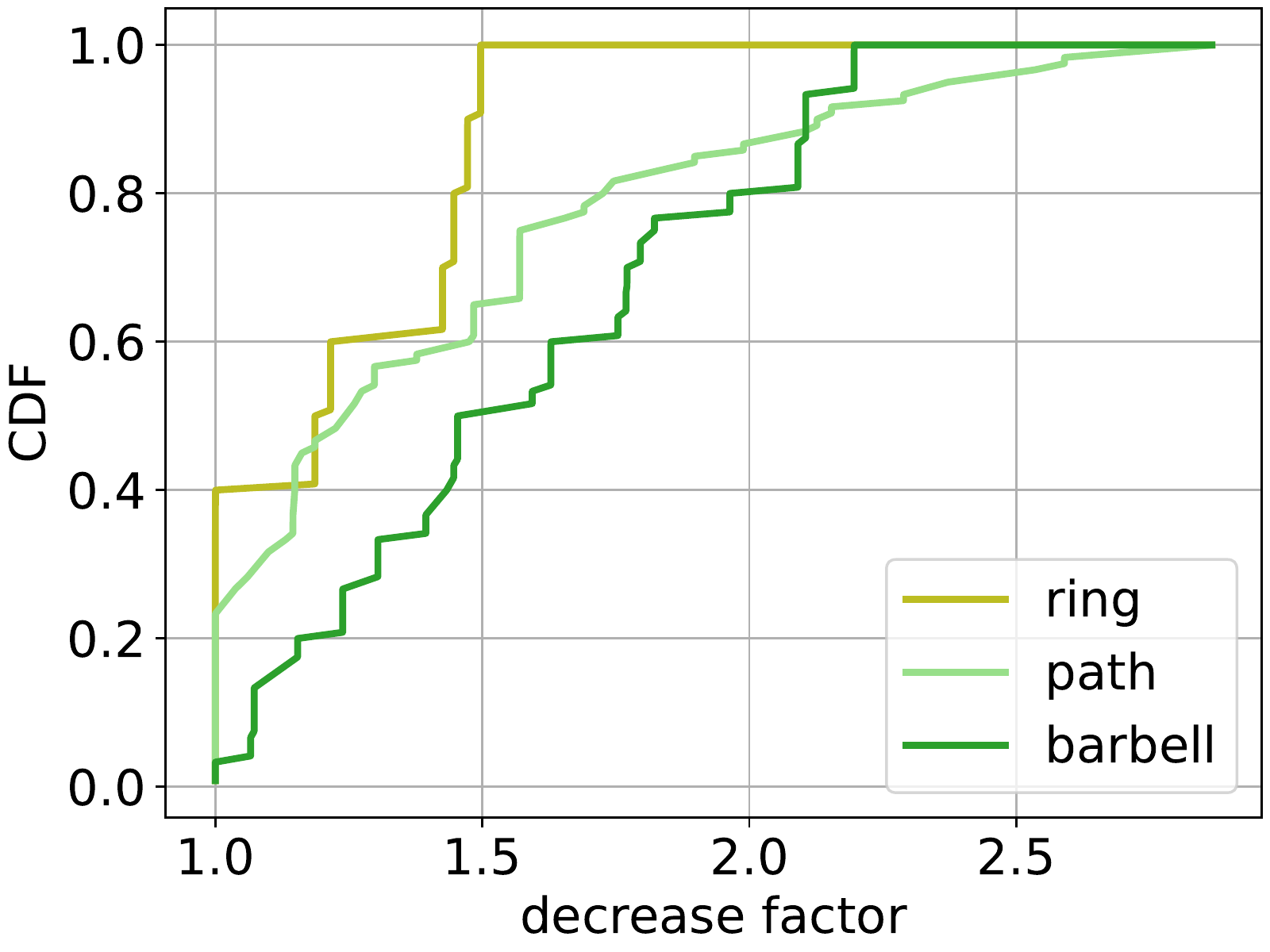}
    \caption{Lateness Gain}
    \label{fig:lateness_gain}
    \end{subfigure}
 	\begin{subfigure}{0.32\textwidth} \centering
    \includegraphics[scale=0.3]{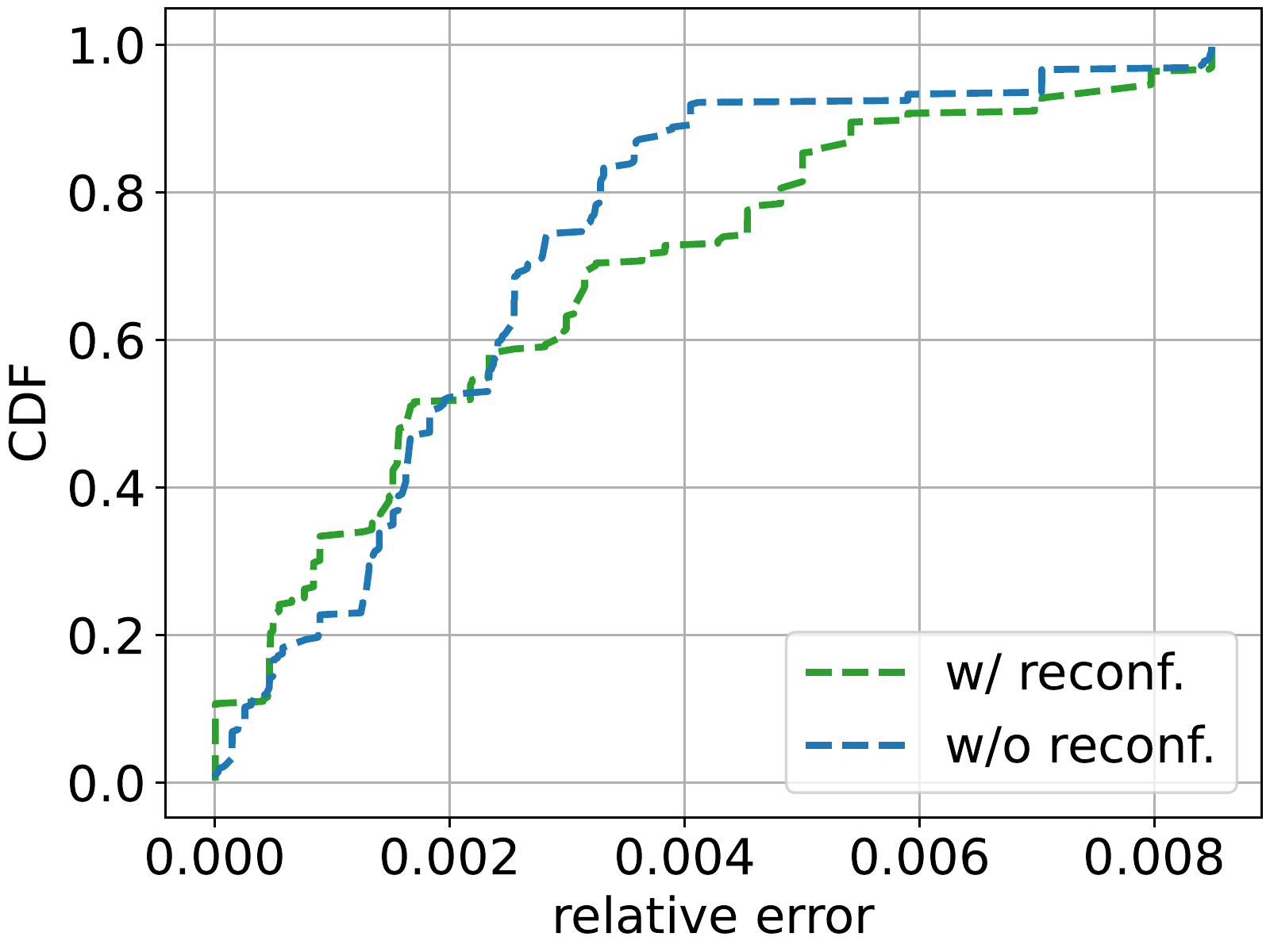}
    \caption{Relative Approximation Error}
    \label{fig:approx_error}
    \end{subfigure}
 	\begin{subfigure}{0.32\textwidth} \centering
    \includegraphics[scale=0.3]{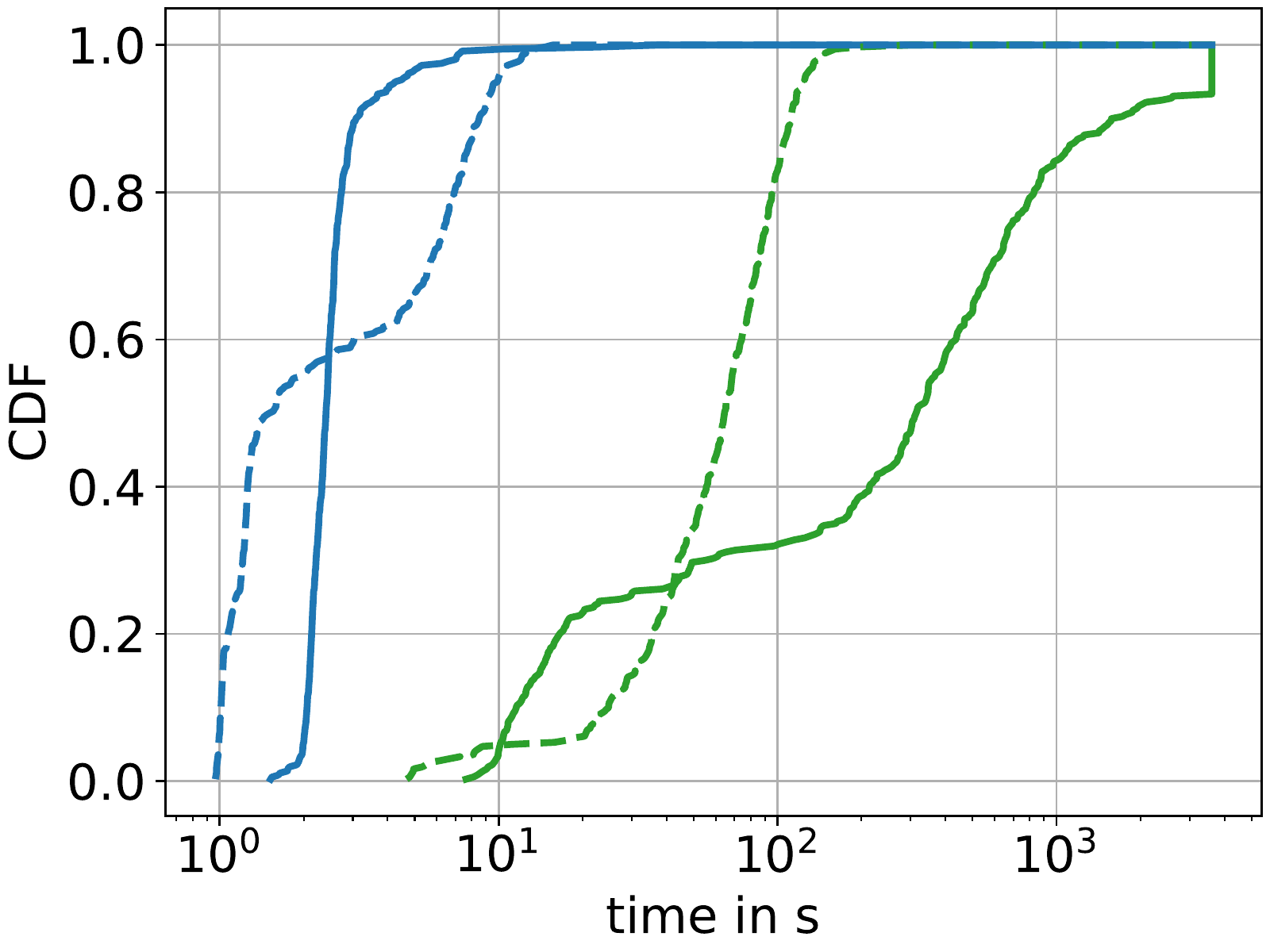}
    \caption{Execution Times}
  	\label{fig:exec_time}
  	\end{subfigure}
  \caption{Results} \label{fig:results}
\end{figure*}

\subsection{Piecewise-Linear Approximation}
For $\Omega \subset \R^2_{++}$ let
\begin{align*}
    g: (0, \infty) \to \R, \, x \mapsto x^{-1}, \\
	h: \Omega \to \R_, \, (x, y) \mapsto g(x) \, y
\end{align*}
and $h(0,0) := 0$. The nonlinear terms in (\ref{forw_del}) and (\ref{exec_del}) are of the form $h(\mu - \lambda, y)$ for $0 \le \lambda \le \mu$ and $y \in \{0, 1\}$. Our goal is to find a piecewise (affine) linear function $\tilde{h}$ for the relaxation $y \in [0,1]$ such that $\tilde{h}$ approximates $h$ for $y \in \{0, 1\}$.

All service rates $\mu$ are bounded from above by some constant $E > 0$: either $\mu$ is the constant forwarding rate $\barmu$ of the transceivers, or $\mu$ is the variable service rate $\mu^n_v$ and due to the vertex capacity constraint (\ref{vertex_util}) bounded by $(c_v - \beta^n)/\alpha^n$. Since $\mu - \lambda$ is not naturally bounded from below, we introduce an artificial lower bound $\eps > 0$ and demand that $\eps < \mu - \lambda$ if $\mu >0$, i.e. we bound the proportional utilization of forwarding and processing capabilities by $\lambda / \mu \in [0, 1 - \eps / \mu]$. Thus allows us to restrict $g$ to the interval $[\eps, E]$ on which the function is bounded.

First, we approximate $g$ by a piece-wise affine linear function $\tildeg$ on the interval $[\eps, E]$ for $\eps \in (0,E)$, see Figure~\ref{fig:approx_g}. Assume we are given a partition $\Pi = \{\pi_k\}_{k \in [K]_0}$ of the interval, i.e. $\pi_0 = \eps$, $\pi_{k-1} < \pi_{k}$ for $k \in [K]$ and $x_K = E$. Any point $x \in [\eps, E]$ can be expressed as a convex combination of two consecutive points in $\Pi$, that is $x = \xi \, \pi_{k-1} + (1 - \xi) \, \pi_{k}$ for appropriate $\xi \in [0,1], k \in [K]$. We approximate $g$ by on $[\pi_{k-1}, \pi_{k}]$ by the linear segment between $g(\pi_{k-1})$ and $g(\pi_{k})$, shifted by a constant $c \ge 0$:
\begin{align*}
	\tildeg(x) &= \xi \, \big(g(\pi_{k-1}) + c \big) + (1 - \xi) \, \big( g(\pi_{k}) + c \big) \\ 
	& =  \xi \, \big( \pi_{k-1}^{-1} + c \big) + (1 - \xi) \, \big( \pi_{k}^{-1} + c \big).
\end{align*}
An algorithmic approach to find a near optimal partition $\Pi$ and constant $c$ for a bounded convex function and given cardinality $K$ can be found in \cite{Rebennack2015-og}. 

Second, we approximate $h$ on $\Omega = \{ \alpha (x, y) \, | \, \alpha \in [0,1], \, x \in [\eps, E], \, y \in [0, 1] \}$ by piece-wise affine linear segments, see Figure~\ref{fig:domain_omega}. Let $\tildePi = \{\tildepi\}_{k \in [K+1]_0} \subset \Omega$ with $\tildepi_k = (\pi_k, 1)$ for $k \in [K]_0$ and $\tildepi_{K+1} = (\pi_K, 0)$. Any point $\omega \in \Omega$ can be expressed as a conic combination of two consecutive points in $\tildePi$, that is  $\omega = \xi_{k-1} \, \tildepi_{k-1} + \xi_{k} \, \tildepi_{k}$ for appropriate $k \in [K+1], \, \boldsymbol{0} \le \boldxi, \; \xi_{k-1} + \xi_{k} \le 1 $. For  $k \in [K]$ we approximate $h(\omega)$ by
\begin{equation}
	\tilde{h}(\omega) = \xi_{k-1} \, \big( \pi_{k-1}^{-1} + c \big) + \xi_{k} \, \big( \pi_{k}^{-1} + c \big),\label{equ:conic_combi_tilde}
\end{equation}
for $k = K + 1$ by
\begin{equation}
	\tilde{h}(\omega) = \xi_K \, (\pi_{K}^{-1} + c). \label{equ:conic_combi_tildeh}
\end{equation}

We add variables $(\xi_k)_{k \in [K+1]_0} \in [0,1]^{K+2}$ for each queue which represent the conic variables (\ref{equ:conic_combi_tilde}), (\ref{equ:conic_combi_tildeh}). Since at most two consecutive entries can be non-zero, $(\xi_k)_{k \in [K+1]_0}$ lies in the special ordered set of type 2 denoted by $\SOS$ \cite{Beale1976-lt}. This membership is implemented by state-of-the-art MILP solvers like Gurobi \cite{gurobi}. For completeness, we state the additional variables and constraints in Appendix \ref{appendix:approx}.

\section{Evaluation}
We evaluate the topology reconfiguration by considering the exact MIQCP and the approximative MILP formulation \textit{with reconfiguration} as stated in Section \ref{section:MIQCP} and \ref{section:approx} and \textit{without reconfiguration} with additional constraints fixingx the substrate topology. We demonstrate that a reconfiguration can reduce the end-to-end delay of services, and that our MILP formulation approximates the original formulation well, while bounding the execution time.

We consider a multicast request with one source, three destinations and a VNF $f$ in between. We assume that the substrate has two vertices for processing with different capacity and that source, destinations and computing vertices are distinct. We evaluate our models on three topologies with six vertices: a path, a barbell and a cycle, see Figure~\ref{fig:topos}. To determine which delay bound can be satisfied for the request, we only consider two of our four objectives: we maximize the number of embedded requests and minimize the lateness. Since we set the requested delay bound to zero, the lateness equals the end-to-end delay of the request. For each topology we evaluated the $6\cdot5\cdot4 = 120$ possibilities of assigning two different processing capacities and the source of the multicast request to three distinct vertices, using the remaining three vertices as the multicast's destinations. The models are described via Pyomo \cite{Pyomo} and solved with Gurobi \cite{gurobi} on a machine with 8 cores and 16GB RAM. Table \ref{table:eval_setup} summarizes the parameters of the setup.

To apply the MILP approximation we need to specify bounds $\mu - \lambda \in [\eps, E]$, where $\mu$ is the service rate and $\lambda$ is the arrival rate of each active forwarding/processing queue, respectively. In our evaluation example, we focus on the lateness of the request and don't minimize the resource consumption. For this reason, the processing capacities of a node are either not, or fully assigned to the request. Since $\lambda \le 3$, we obtain bounds $[1, 4]$ for the forwarding queues, $[2, 5]$ for the processing vertex with smaller capacity and $[47, 50]$ for the vertex with larger capacity. We choose six base points for $[1, 4]$, four base points for for $[2, 5]$ and 2 base points (i.e. just the upper and the lower bound) for $[47, 50]$ as these choices lead to an approximation error smaller than $0.01$ per queue.

The results of our evaluation are shown in Figure~\ref{fig:results}. As we expected, the reconfiguration of the lightpath topology can decrease the end-to-end delay: here, by a factor up to 2.8. However, the gain depends on the substrate topology, available transceivers on the vertices and the distribution of ingresses, egresses \& processing capabilities. The MILP formulation approximate the exact MIQCP well; in 80\% of our evaluated scenarios the proportion of the absolute difference of the computed minimal latenesses and the exact MIQCP lateness was smaller than 0.01, see Figure~\ref{fig:approx_error}. Moreover, the MILP approximation bounded the execution time by $\sim 100$s. In contrast, solving the MIQCP took in 15\% of our evaluated scenarios more than $1000$s and exceeded our processing bound of 1h in more than 5\%. Nevertheless, the approximation could not always decrease the execution time, compare Figure~\ref{fig:exec_time}.

\section{Summary and Outlook}

We showed that a reconfiguration of the lightpath topology in the context of NFV can decrease the end-to-end delay of services. Solving the large optimization problems in practice will be challenging. However, our work can be a base for reducing the model in order to obtain smaller MIQCP/MILP instances or serve as a baseline for heuristics.

\printbibliography

\appendix

For completeness and as a reference for implementation we add the details of the MILP approximation. 

\subsection{Wavelength Assignment Only Formulation} \label{appendix:WA only}

For completeness, we describe our adoption of the problem.
For all $w, w^\prime \in \calV$ we denote the unique given path from $w$ to $w^\prime$ with length $J$ by $p_{w, w^\prime} \in \calV^{J + 1}$ and its associated propagation delay by $d_{w, w^\prime} \ge 0$. After deciding which lightpath to establish, we no longer have to decide the routing, but are still left with the wavelength assignment; instead of considering binary variables $l_{w, w^\prime, e, \gamma} \in \{0, 1\}$ we consider $l_{w, w^\prime, \gamma} \in \{0, 1\}$ to indicate a lightpath establishment from $w$ to $w^\prime$ on $\gamma \in \Gamma$. Here, we express that $e \in \calE$ is used in path $p_{w, w^\prime}$ by writing $e \sqsubset p_{w, w^\prime}$.
\begin{flalign*}
\shortintertext{$\forall w, w^\prime \in \calV:$}
&& \sum_{r \in \calR} \, \sum_{a \in A} \, \sum_{\substack{v, v^\prime \in \calV \\ v \neq v^\prime}}  \lambda^a_{v, v^\prime, w, w^\prime} &\le \sum_{\gamma \in \Gamma} \barmu \cdot  l_{w, w^\prime, \gamma} \\
&& \sum_{\gamma \in \Gamma} l_{w, w^\prime, \gamma} &\le 1 \\
\shortintertext{$\forall e \in \calE, \, \gamma \in \Gamma:$}
&& \sum_{\substack{w, w^\prime \in \calV \\ e \sqsubset p_{w, w^\prime}}} l_{w, w^\prime, \gamma} &\le 1 \\
\shortintertext{$\forall w, \, w^\prime \in \calV, \, \gamma \in \Gamma:$}
&& l_{w, w^\prime, \gamma} &= l_{w^\prime, w, \gamma}  \\
\shortintertext{$\forall w \in \calV:$}
&& \sum_{w^\prime \in \calV} \sum_{\gamma \in \Gamma} l_{w, w^\prime, \gamma} &\le \deg (w) 
\end{flalign*}
The propagation delays (\ref{prop_del}) in the delay constraints reduce to the linear term
\begin{equation*}
	\sum_{j \in [J-1]} \sum_{w, w^\prime \in \calV} d_{w, w^\prime} \cdot z^{n_j n_{j+1}}_{v_j, v_{j+1}, w, w^\prime}.
\end{equation*}
In the objective function we set
\begin{equation*}
	o_{\text{path}} = \sum_{w, w^\prime \in \calV} \, \sum_{\gamma \in \Gamma} d_{w, w^\prime} \cdot l_{w, w^\prime, \gamma}.
\end{equation*}

\subsection{Piecewise Linear Approximation} \label{appendix:approx}

For simplicity, we assume that the lower and upper bounds $0 <\eps < E$ and the partitions $\Pi \subset [\eps, E]$ containing $K+1$ base points are the same for all queuing terms. It is straight forward to use different parameters for every queue. In that case, some caution is needed to deal with vertices where $E <= \eps$, e.g. for vertices with no processing capacity at all.

We replace the utilization constraints (\ref{capacity_constr2}), (\ref{lightpath_activation_with_bitrate}) by
\begin{flalign}
\shortintertext{$\forall \, v \in \calV \, n \in N_{\calF}:$}
&& \eps \cdot y^n_v &\le \mu^n_v - \sum_{n^{\minus} \in N^-(n)} \, \sum_{v^\prime, w \in \calV} \lambda^{n^{\minus}n}_{v^\prime, v, w, v} \label{equ:eps_node_capacity_constr} \\
\shortintertext{$\forall \, w, w^\prime \in \calV, \, w \neq w^\prime:$}
&& \eps \, \sum_{\gamma \in \Gamma} l_{w, w^\prime, \gamma} &\le \sum_{\gamma \in \Gamma} \barmu \cdot l_{w, w^\prime, \gamma}  - \sum_{r \in \calR} \, \sum_{a \in A} \, \sum_{\substack{v, v^\prime \in \calV \\ v \neq v^\prime}} \lambda^a_{v, v^\prime, w, w^\prime},\label{equ:eps_link_capacity_constr} 
\end{flalign}
set $\pi_K = \pi_{K+1}$ and add the the constraints
\begin{flalign}
\shortintertext{$\forall \, n \in N_{\calF}, \, a \in A, \,  v, v^\prime, w, w^\prime \in \calV, \, k \in [K+1]_0 :$}
&& \xi^n_{v, k}, \, \xi^a_{v, v^\prime, w, w^\prime, k} & \in [0,1] \\
&& (\xi^n_{v, k})_{k \in [K+1]_0} & \in \SOS \\
&& (\xi^a_{v, v^\prime, w, w^\prime, k})_{k \in [K+1]_0} & \in \SOS \\
\shortintertext{$\forall \, n \in N_{\calF}, \, v \in \calV:$}
&& y^n_v & = \sum_{k\in [K]_0} \xi^n_{v, k} \\
&& \mu^{n}_{v} -  \sum_{n^{\minus} \in N^-(n)} \, \sum_{v^\prime, w \in \calV} \lambda^{n^{\minus}n}_{v^\prime, v, w, v} & = \sum_{k\in [K+1]_0} \pi_k \cdot \xi^n_{v, k}  \nonumber \\
\shortintertext{$\forall \, a \in A, \, v, v^\prime, w, w^\prime \in \calV:$}
&& z^a_{v, v^\prime, w, w^\prime} &= \sum_{k\in [K]_0} \xi^a_{v, v^\prime, w, w^\prime, k} \\
&& \barmu - \sum_{r^\prime \in \calR} \, \sum_{a^\prime \in A} \, \sum_{u, u^\prime \in \calV} \lambda^{a^\prime}_{u, u^\prime, w, w^\prime} &=  \sum_{k\in [K+1]_0} \pi_k \cdot \xi^a_{v, v^\prime, w, w^\prime, k}.
\end{flalign}
The sum of the forwarding and processing delays (\ref{forw_del}), (\ref{exec_del}) is then approximated by 
\begin{equation*}
\begin{split}
 \sum_{j \in [J-1]} \, \sum_{\substack{w, w^\prime \in \calV \\ w \neq w^\prime}} \, \sum_{k \in [K]_0} \, \big(\pi_k^{-1} + c \big) \cdot \xi^{n_{j}n_{j+1}}_{v_j, v_{j+1}, w, w^\prime, k} 
 \\ + \sum_{j \in [J-2]} \, \sum_{k \in [K]_0} \big( \pi_k^{-1} + c \big) \cdot \xi^{n_{j+1}}_{v_{j+1}, k}.
 \end{split}
 \end{equation*}

\end{document}

%% file: graphics/motivation1.tex
\begin{tikzpicture}[scale=0.5]
\tikzstyle{every node}+=[inner sep=0pt]
\draw [black, very thick] (1,-1) circle (1); \draw [gray] (1, -1) node {$v_1$}; 
\draw [black, very thick] (1,-5) circle (1); \draw [gray] (1, -5) node {$v_2$}; 
\draw [black, very thick] (3.303, -2.303) rectangle (4.707, -3.707); \draw [gray] (4, -3) node {$v_3$}; 
\draw [black, very thick] (6.303,-1.808) rectangle (8.707, -4.202); \draw [gray] (7.505, -3) node {$v_4$};	 
\draw [black, very thick] (11,-1) circle (1); \draw [gray] (11, -1) node {$v_5$}; 
\draw [black, very thick] (11,-5) circle (1); \draw [gray] (11, -5) node {$v_6$}; 
\draw [black, dashed] (1.924,-1.383) -- (3.303,-2.303); 
\draw [black, dashed] (1.924,-4.617) -- (3.303,-3.707); 
\draw [black, dashed] (4.707,-3) -- (6.303,-3); 
\draw [black, dashed] (8.707,-2.303) -- (10.076,-1.383); 
\draw [black, dashed] (8.707,-3.707) -- (10.076,-4.707); 
\end{tikzpicture}

%% file: graphics/motivation2.tex
\begin{tikzpicture}[scale=0.5]
\tikzstyle{every node}+=[inner sep=0pt]
\draw [blue, thick] (1.924,-1.383) -- (3.303,-2.303); 
\draw [blue, thick] (1.924,-4.617) -- (3.303,-3.707); 
\draw [blue, thick] (4.707,-3) -- (6.303,-3); 
\draw [blue, thick] (8.707,-2.303) -- (10.076,-1.383); 
\draw [blue, thick] (8.707,-3.707) -- (10.076,-4.707); 
\draw [black, very thick] (1,-1) circle (1);  
\draw [black, very thick] (1,-5) circle (1);  
\draw [black, very thick] (3.303, -2.303) rectangle (4.707, -3.707); 
\draw [black, very thick] (6.303,-1.808) rectangle (8.707, -4.202); \draw (7.505, -3) node {$f, g$};	 
\draw [black, very thick] (11,-1) circle (1);  
\draw [black, very thick] (11,-5) circle (1);  
\end{tikzpicture}

%% file: graphics/motivation3.tex
\begin{tikzpicture}[scale=0.5]
\tikzstyle{every node}+=[inner sep=0pt]
\draw [blue, thick] (1.924,-1.383) -- (3.303,-2.303); 
\draw [red, thick] (1.924,-4.617) -- (3.303,-3.707); 
\draw [red, thick] (3.303, -3.707) -- (4.707, -3.1); 
\draw [blue, thick] (4.707,-3) -- (6.303,-3); 
\draw [red, thick] (4.707,-3.1) -- (6.303,-3.1); 
\draw [blue, thick] (6.303, -3) -- (8.707, -2.303); %
\draw [blue, thick] (8.707,-2.303) -- (10.076,-1.383); 
\draw [blue, thick] (8.707,-3.707) -- (10.076,-4.707); 
\draw [black, very thick] (1,-1) circle (1);  
\draw [black, very thick] (1,-5) circle (1);  
\draw [black, very thick] (3.303, -2.303) rectangle (4.707, -3.707); \draw (4, -2.9) node {$f$}; 
\draw [black, very thick] (6.303,-1.808) rectangle (8.707, -4.202); \draw (7.505, -3.1) node {$g$};	 
\draw [black, very thick] (11,-1) circle (1);  
\draw [black, very thick] (11,-5) circle (1);  
\end{tikzpicture}

%% file: graphics/simple_nfc_tikz.tex
\begin{tikzpicture}[scale=0.2]
\tikzstyle{every node}+=[inner sep=0pt]
\draw [black] (1,-1) circle (1);
\draw (1,-1) node {$s$};
\draw [black] (5,-1) circle (1);
\draw (5,-1) node {$f$};
\draw [black] (9,-1) circle (1);
\draw (9,-1) node {$d$};
\draw [black, ->] (2,-1) -- (4,-1);
\draw [black, ->] (6,-1) -- (8,-1);
\end{tikzpicture}

%% file: graphics/path_graph.tex
\begin{tikzpicture}[scale=0.2]
\tikzstyle{every node}+=[inner sep=0pt]
\draw [black] (1,-1) circle (1);
\draw [black] (5,-1) circle (1);
\draw [black] (9,-1) circle (1);
\draw [black] (13,-1) circle (1);
\draw [black] (17,-1) circle (1);
\draw [black] (21,-1) circle (1);
\draw [black] (2,-1) -- (4,-1);
\draw [black] (6,-1) -- (8,-1);
\draw [black] (10,-1) -- (12,-1);
\draw [black] (14,-1) -- (16,-1);
\draw [black] (18,-1) -- (20,-1);
\end{tikzpicture}

%% file: graphics/barbell_graph.tex
\begin{tikzpicture}[scale=0.2]
\tikzstyle{every node}+=[inner sep=0pt]
\draw [black] (1,-1) circle (1); 
\draw [black] (1,-5) circle (1); 
\draw [black] (4,-3) circle (1); 
\draw [black] (8,-3) circle (1);	 
\draw [black] (11,-1) circle (1); 
\draw [black] (11,-5) circle (1); 
\draw [black] (1,-2) -- (1,-4); 
\draw [black] (1.924,-1.383) -- (3.303,-2.303); 
\draw [black] (1.924,-4.617) -- (3.303,-3.707); 
\draw [black] (5,-3) -- (7,-3); 
\draw [black] (8.707,-2.303) -- (10.076,-1.383); 
\draw [black] (8.707,-3.707) -- (10.076,-4.707); 
\draw [black] (11,-2) -- (11,-4); 
\end{tikzpicture}

%% file: graphics/ring_graph.tex
\begin{tikzpicture}[scale=0.2]
\tikzstyle{every node}+=[inner sep=0pt]
\draw [black] (1,-3) circle (1);
\draw [black] (4,-1) circle (1);
\draw [black] (4,-5) circle (1);
\draw [black] (8,-1) circle (1);
\draw [black] (8,-5) circle (1);
\draw [black] (11,-3) circle (1);
\draw [black] (1.707,-2.303) -- (3.076,-1.383);
\draw [black] (1.707,-3.707) -- (3.076,-4.627);
\draw [black] (5,-1) -- (7,-1);
\draw [black] (5,-5) -- (7,-5);
\draw [black] (8.924,-1.282) -- (10.303,-2.303);
\draw [black] (8.924,-4.707) -- (10.303,-3.707);
\end{tikzpicture}